\documentclass[12pt]{article}
\pdfoutput=1
\usepackage{amssymb,amsmath}
\usepackage{graphicx}
\usepackage{color}
\usepackage[colorlinks=true
,urlcolor=blue
,citecolor=blue
,linkcolor=blue
,pagecolor=blue
,linktocpage=true
,pdfproducer=medialab
]{hyperref}
\usepackage[a4paper,width=17.1cm]{geometry}
\makeatletter \renewcommand{\@dotsep}{10000} \makeatother
\usepackage{appendix}
\usepackage{subcaption}
\usepackage{mwe}

\newcommand{\gmu}{\ensuremath{(g-2)_{\mu}}}
\newcommand{\damu}{\ensuremath{\Delta a_{\mu}}}

\begin{document}

\begin{center}

 {\Large  \textbf{SU(5) with Non-Universal Gaugino Masses } 
 } \vspace{1cm}

{  M. Adeel Ajaib\footnote{ E-mail: adeel@udel.edu}
 } \vspace{.5cm}

{ \it 
Department of Mathematics, Statistics and Physics,\\ Qatar University, Doha, Qatar \\
} \vspace{.5cm}

\vspace{1.5cm}
 {\bf Abstract}\end{center}

We explore the sparticle spectroscopy of the supersymmetric SU(5) model with non-universal gaugino masses in light of latest experimental searches. We assume that the gaugino mass parameters are independent at the GUT scale. We find that the observed deviation in the anomalous magnetic moment of the muon can be explained in this model. The parameter space that explains this deviation predicts a heavy colored sparticle spectrum whereas the sleptons can be light. We also find a notable region of the parameter space that yields the desired relic abundance for dark matter. In addition, we analyze the model in light of latest limits from direct detection experiments and find that the parameter space corresponding to the observed deviation in the muon anomalous magnetic moment can be probed at some of the future direct detection experiments.

\newpage

\renewcommand{\thefootnote}{\arabic{footnote}}
\setcounter{footnote}{0}

\section{Introduction}

Extensive search for Supersymmetry (SUSY) continues at various fronts such as the Large Hadron Collider (LHC) and direct/indirect detection experiments. The elegance of SUSY lies in the fact that it leads to gauge coupling unification and provides a viable candidate for cold dark matter (the neutralino) \cite{Jungman:1995df}. The discovery of a 125 GeV Higgs boson, although consistent with predictions from the Minimal Supersymmetric Standard Model (MSSM), implies severe constraints on the parameter space of SUSY. 
The ATLAS and CMS experiments at 13 TeV LHC (with an integrated luminosity of 36 fb$^{-1}$) have recently reported updated bounds on various sparticle masses. For instance,
the reported limit on the first/second generation squark masses from LHC is $m_{\tilde{q}} \simeq 1.6$ TeV \cite{lhc-squark}. The currrent limits on the gluino mass is $m_{\tilde{g}} \simeq 1.9$ TeV and the stop mass is $m_{\tilde{t}} \simeq 1$ TeV \cite{lhc-stop}. In addition, current searches for the charginos have not resulted in any signals and the present limit on its mass is $m_{\tilde{\chi}^\pm} \simeq 430$ GeV  \cite{lhc-chargino}. The High Luminosity LHC (HL-LHC) is expected to improve these limits if no SUSY signals are found ~\cite{upgrade,gershtein,atlaswiki,Baer:2017pba}.

Another possible signature of SUSY may be the observed deviation in the muon anomalous magnetic moment $a_{\mu}=(g-2)_{\mu}/2$ (muon $g-2$) from its SM prediction 
 \cite{Hagiwara:2011af}
\begin{eqnarray}
\label{gg-22}
\Delta a_{\mu}\equiv a_{\mu}({\rm exp})-a_{\mu}({\rm SM})= (28.6 \pm 8.0) \times 10^{-10}.
\end{eqnarray}
In our analysis we show that the SU(5) model with non-universal gaugino masses can explain the above deviation in $\gmu$.

The paper is orgranized as follows: In section \ref{g-2}, we briefly review the SUSY contribution to the muon anomalous magnetic moment and present the expression for $\damu$. Section \ref{sec:parameter} describes our scanning procedure, the constraints we implement and the parameter space of the SU(5) model we explore. In section \ref{sec:results}, we present the results of our parameter space scan. We conclude in section \ref{sec:conclude}

\section{\label{g-2}The Muon Anomalous Magnetic Moment}

The leading  contribution from low scale supersymmetry  to the muon anomalous magnetic moment is given by \cite{Moroi:1995yh, Martin:2001st}:

\begin{eqnarray}
\label{eqq1}
\Delta a_\mu &=& \frac{\alpha \, m^2_\mu \, \mu\,  \tan\beta}{4\pi} {\bigg \{ }
\frac{M_{2}}{ \sin^2\theta_W \, m_{\tilde{\mu}_{L}}^2}
\left[ \frac{f_{\chi}(M_{2}^2/m_{\tilde{\mu}_{L}}^2)-f_{\chi}(\mu^2/m_{\tilde{\mu}_{L}}^2)}{M_2^2-\mu^2} \right] 
\nonumber\\
&+&
\frac{M_{1} }{ \cos^2\theta_W \, (m_{\tilde{\mu}_{R}}^2 - m_{\tilde{\mu}_{L}}^2)}
\left[\frac{f_{N}(M^2_1/m_{\tilde{\mu}_{R}}^2)}{m_{\tilde{\mu}_{R}}^2} - \frac{f_{N}(M^2_1/m_{\tilde{\mu}_{L}}^2)}{m_{\tilde{\mu}_{L}}^2}\right] \, {\bigg \} },
\end{eqnarray}
where $\alpha$ is the fine-structure constant, $\mu$ is the bilinear Higgs mixing term, $m_\mu$ is the muon mass, and $\tan\beta$ is the ratio of the vacuum expectation values (VEV) of the MSSM Higgs doublets. $M_1$ and $M_2$ denote the $U(1)_Y$ and $SU(2)$ gaugino masses respectively, $\theta_W$  is the weak mixing angle, and $m_{\tilde{\mu}_{L}}$ and $m_{\tilde{\mu}_{R}}$ are the left and right handed smuon masses. The loop functions are defined as follows:
\begin{eqnarray}
f_{\chi}(x) &=& \frac{x^2 - 4x + 3 + 2\ln x}{(1-x)^3}~,\qquad ~f_{\chi}(1)=-2/3, \\
f_{N}(x) &=& \frac{ x^2 -1- 2x\ln x}{(1-x)^3}\,,\qquad\qquad f_{N}(1) = -1/3 \, .
\label{eqq2}
\end{eqnarray}
The first term in equation (\ref{eqq1}) stands for the dominant contribution from one loop diagram with charginos (Higgsinos and Winos), while the second term entails contributions from the bino-smuon loop.

\section{Non-Universal Gaugino Masses in SU(5)}\label{sec:parameter}

In the $SU(5)$ GUT, the  SM fermions of each family are allocated to the following representations: $\bar{5} \supset  (d^c, L)$ and $10\supset (Q, u^c, e^c)$.
We consider two independent Soft SUSY Breaking (SSB) scalar mass terms at $M_{\rm GUT}$, namely, $m_{\bar{5}}$ and $m_{10}$, for the matter multiplets.
For simplicity, we will assume that at the GUT  scale  we have  $m_{\bar{5}} = m_{H_u} = m_{H_d}$, where  $m_{H_u}$ and $m_{H_d}$ are the mass parameters of the MSSM Higgs doublets, which belong to the $5 (H_u)$ and $\bar{5} (H_{d})$ representations of $SU(5)$  ~\cite{Profumo:2003ema,Gogoladze:2008dk}.
Therefore, in the SU(5) scenario the  SSB masses at $M_{\rm GUT}$ are as follows:
\begin{eqnarray}
 &&  m_{\tilde{D}^c} =  m_{\tilde{L}} = m_{H_u} = m_{H_d} = m_{\bar 5},
\nonumber \\
 &&  m_{\tilde{Q}} = m_{\tilde{U}^c} = m_{\tilde{E}^c} = m_{10} ,
\label{BC}
\end{eqnarray}
Non-universality of gaugino masses have been considered in several studies \cite{Kawamura:2017amp} and many have made attempts to explain the observed deviation in $\gmu$ in this context \cite{Akula:2013ioa, Gogoladze:2014cha, Gogoladze:2016jvm}. For example, it was shown in \cite{Gogoladze:2014cha} that the $g-2$ anomaly can be resolved by employing non-universal gaugino masses in SUSY $SO(10)$. Furthermore, it was shown in \cite{Ajaib:2014ana} that the resolution of the muon $g-2$ anomaly is compatible with a 125 GeV Higgs boson mass, the WMAP relic dark matter density and excellent $t$-$b$-$\tau$ Yukawa unification.

It has also been pointed out that non-universal MSSM gaugino masses at $ M_{\rm GUT} $ can arise from non-singlet F-terms, compatible with the underlying GUT symmetry such as $SU(5)$ and $SO(10)$ \cite{Martin:2009ad}. Non-universal gauginos can also be generated  from an $F$-term which is a linear combination of two distinct fields of different dimensions \cite{Martin:2013aha}.   One can also consider two distinct sources for supersymmetry breaking \cite{Anandakrishnan:2013cwa}. {With many distinct} possibilities available for realizing nonuniversal gaugino masses we employ three independent masses for the  MSSM  gauginos in SUSY $SU(5)$ GUT.

{We employ Isajet~7.84 \cite{ISAJET} interfaced with Micromegas 2.4 \cite{Belanger:2008sj} to perform random scans over the parameter space.} We use Micromegas to calculate the relic density and $BR(b \rightarrow s \gamma)$. The function RNORMX \cite{Leva} is employed
to generate a Gaussian distribution around random points in the parameter space. Further details regarding our scanning procedure can be found in \cite{Ajaib:2015ika}.  After collecting the data, we impose the following experimental constraints on the parameter space:

\begin{table}[h!]\centering
\begin{tabular}{rlc}
$123~{\rm GeV} \leq  m_h  $ & $\leq 127~{\rm GeV}$~~
\\
$ 0.8 \times 10^{-9} \leq BR(B_s \rightarrow \mu^+ \mu^-) $&$ \leq\, 6.2 \times 10^{-9} \;
 (2\sigma)$        &        \\
$2.99 \times 10^{-4} \leq BR(b \rightarrow s \gamma) $&$ \leq\, 3.87 \times 10^{-4} \;
 (2\sigma)$ &     \\
$0.15 \leq \frac{BR(B_u\rightarrow
\tau \nu_{\tau})_{\rm MSSM}}{BR(B_u\rightarrow \tau \nu_{\tau})_{\rm SM}}$&$ \leq\, 2.41 \;
(3\sigma)$. &  \\
 $m_{\tilde{g}}$ &$ \gtrsim  1.9~{\rm TeV}$ \\
  $m_{\tilde{t}}$ &$ \gtrsim  1 ~{\rm TeV}$\\
   $m_{\tilde{\chi}^\pm} $ & $ \gtrsim  430~{\rm GeV}$\\
  $ 20.6 \times 10^{-10} <  \Delta a_\mu  $ &$< 36.6 \times 10^{-10} \,\,  (1\sigma)$ 

\end{tabular}\label{table}
\end{table}
The ranges of the parameters for this model are as follows:
\begin{eqnarray}
 0\leq & m_{\bar 5} & \leq 30\, \textrm{TeV} \nonumber \\
 0\leq & m_{10} & \leq 30 \, \textrm{TeV}  \nonumber \\
 -3 \le & A_0/m_{\bar 5} & \le 3 \nonumber \nonumber \\
 2 \le & \tan\beta &  \le 60 \nonumber \\
 0 \le & M_1 &\le {\mathrm {5 \ TeV} } \nonumber \\
0 \le & M_2 &\le {\mathrm {5 \ TeV} } \nonumber \\
{\mathrm {-5 \ TeV} } \le & M_3 &\le 0 \nonumber  \\
 &\mu >0 \nonumber 
\end{eqnarray}
Here  $M_{1}$, $M_{2}$, and $M_{3}$ denote the SSB gaugino masses for $U(1)_{Y}$, $SU(2)_{L}$ and $SU(3)_{c}$ respectively. $ \tan\beta $ is the ratio of the vacuum expectation values (VEVs) of the two MSSM Higgs doublets, and $ A_{0} $ is the universal SSB trilinear scalar interaction (with corresponding Yukawa coupling factored out).   In order to obtain the correct sign for the desired contribution to $ \gmu $, we set same signs for the parameters $\mu$, $M_1$ and $M_2$. We choose $m_t = 173.3\, {\rm GeV}$.

\section{Results and Analysis}\label{sec:results}

In this section we present our results for the parameter space scan described in section \ref{sec:parameter}. 
In Figures 1-6, \textit{green} points satisfy the sparticle mass constraints and $B$-physics constraints described in section \ref{sec:parameter}. \textit{Brown} points form a subset of the \textit{green} points and satisfy $0.001 \le \Omega h^2 \le 1 $. We choose a wider range for the relic density due to the uncertainties involved in the numerical calculations of various spectrum calculators. Moreover, dedicated scans within the \textit{brown} regions can yield points compatible with the current WMAP range for relic abundance. 
\textit{Orange} points form a subset of the \textit{green} points and satisfy the muon $g-2$ constraint presented in section  \ref{sec:parameter}. From the figures we can observe that there is a considerable region of the parameter space that satisfies the sparticle mass and $B$-physics constraints. In addition, there is a notable region of the parameter space that satisfies the desired relic density constraint (\textit{brown} points) and  the $\gmu$ constraint (\textit{orange} points).

\begin{figure}
        \includegraphics[width=8cm, height=6.5cm]{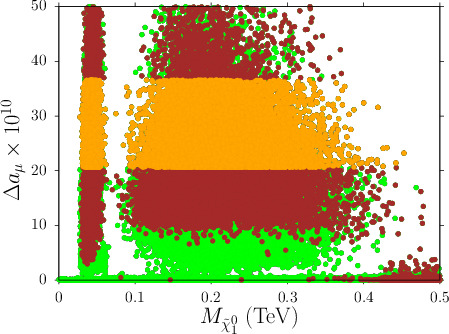}\hfill
        \includegraphics[width=8cm, height=6.5cm]{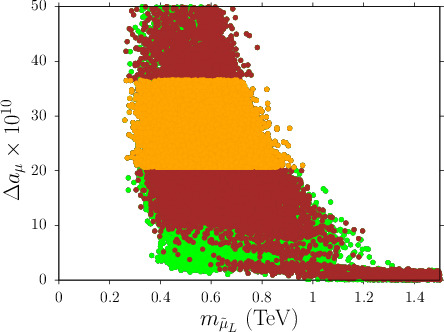}\vspace{5mm} \\ 
        \includegraphics[width=8cm, height=6.5cm]{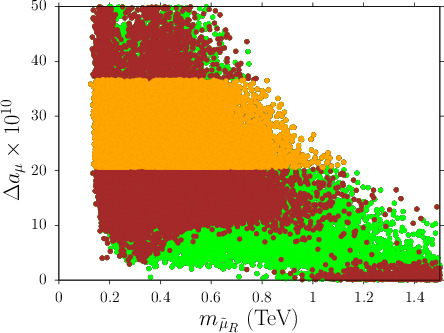}\hfill
        \includegraphics[width=8cm, height=6.5cm]{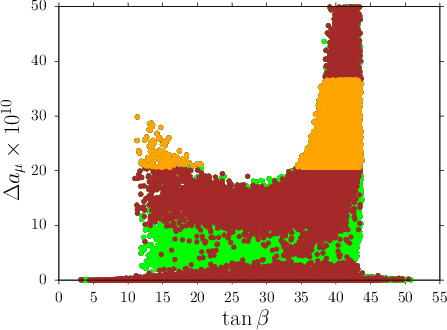}\vspace{5mm} \\ 
         \includegraphics[width=8cm, height=6.5cm]{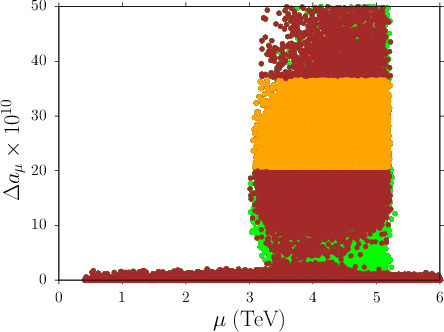}\hfill
         \includegraphics[width=8cm, height=6.5cm]{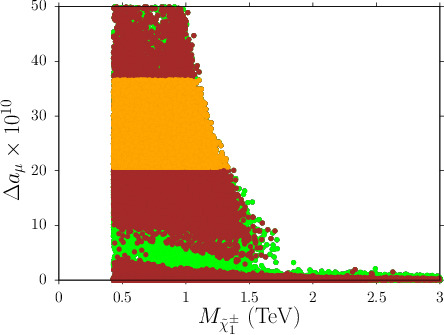}
\caption{Plots in the $\damu-M_{\tilde{\chi}_{1}^{0}}$, $\damu-m_{\tilde{\mu}_{L}}$, $\damu-m_{\tilde{\mu}_{R}}$,  $\damu-\tan\beta$, $\damu-\mu$, and $\damu-M_{\tilde{\chi}_1^\pm}$ planes. \textit{Green} points satisfy the sparticle mass constraints and B-physics constraints described in Section \ref{sec:parameter}. \textit{Brown} points form a subset of the \textit{green} points and satisfy $0.001 \le \Omega h^2 \le 1 $. \textit{Orange} points are subset of the green points and satisfy the muon $g-2$ constraint described in Section  \ref{sec:parameter}.}
\label{fig:damu-plots}
\end{figure}

In Figure, \ref{fig:damu-plots} we display our results in the $\damu-M_{\tilde{\chi}_{1}^{0}}$, $\damu-m_{\tilde{\mu}_{L}}$, $\damu-m_{\tilde{\mu}_{R}}$,  $\damu-\tan\beta$, $\damu-\mu$, and $\damu-M_{\tilde{\chi}_1^\pm}$ planes.  We can see that the resolution of the $\gmu$ anomaly implies an upper bound of $\sim$ 400 GeV on the neutralino mass. In addition, the following bounds can be deduced for the left and right handed smuon masses:  
$300 {\rm \ GeV} \lesssim m_{\tilde{\mu}_{L}} \lesssim 1 {\rm \ TeV}$ and
$150 {\rm \ GeV} \lesssim m_{\tilde{\mu}_{R}} \lesssim 1.2 {\rm \ TeV}$.
 Moreover, good $\gmu$ implies fairly large values of the parameter $\mu$, i.e., $3 {\rm \ TeV} \lesssim \mu \lesssim 5.5 {\rm \ TeV}$ and $\tan\beta$ is restricted to the following intervals:
 $35 \lesssim \tan\beta \lesssim 45$ and
 $12 \lesssim \tan\beta \lesssim 20$. The lower right panel shows that the chargino mass has an upper bound of around 1.5 TeV due to the $\gmu$ constraint.
  Note that earlier studies have found that there are several factors which can lead to a large SUSY contribution to $\gmu$. These include the case when $M_1$, $M_2$ and $\mu$ have the same sign~\cite{Pokoroski}, in which case both of the terms arising from chargino-sneutrino and bino-smuon loops in equation (\ref{eqq1}) will be positive. Furthermore, large values of $\mu\tan\beta$ and light smuons can also lead to large $\Delta a_\mu$  as can be seen from equation (\ref{eqq1}). Results displayed in Figure \ref{fig:damu-plots} are consistent with these observations.

\begin{figure}
        \includegraphics[width=8cm, height=6.5cm]{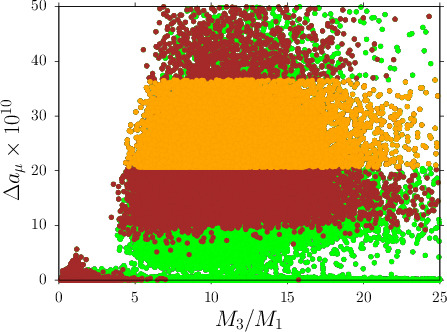}\hfill
        \includegraphics[width=8cm, height=6.5cm]{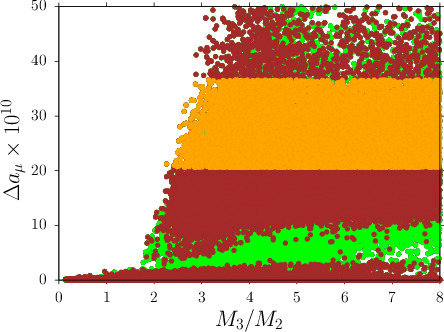}\vspace{2mm} \\ 
        \includegraphics[width=8cm, height=6.5cm]{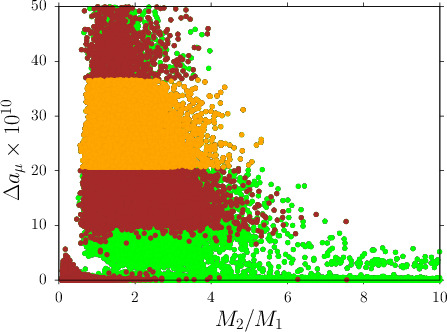}\hfill
        \includegraphics[width=8cm, height=6.5cm]{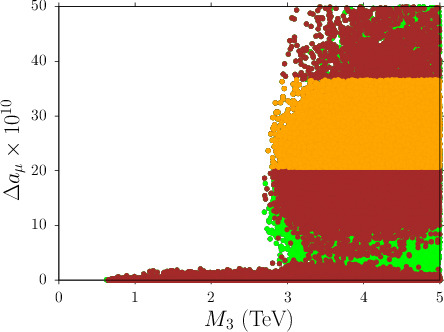}\vspace{2mm} \\ 
         \includegraphics[width=8cm, height=6.5cm]{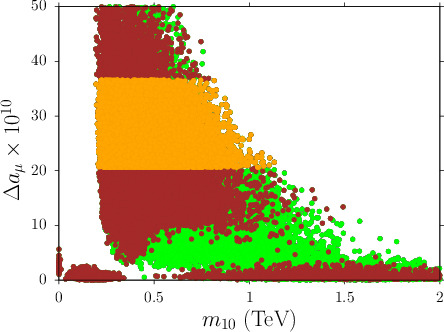}\hfill
         \includegraphics[width=8cm, height=6.5cm]{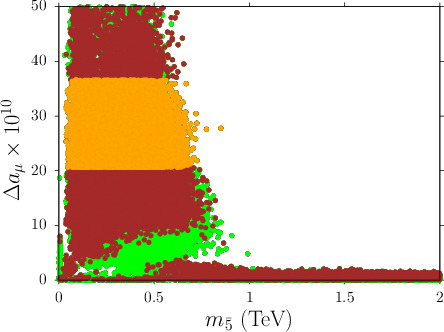}
\caption{Plots in the $\damu-M_{3}/M_{1}$, $\damu-M_{3}/M_{2}$, $\damu-M_{2}/M_{1}$, $\damu-M_{2}$, $\damu-M_{3}$ and $\damu-m_{16}$ planes. Color coding is the same as Figure \ref{fig:damu-plots}.}
\label{fig:damu-plots2}
\end{figure}

\begin{figure}
        \includegraphics[width=8cm, height=6.5cm]{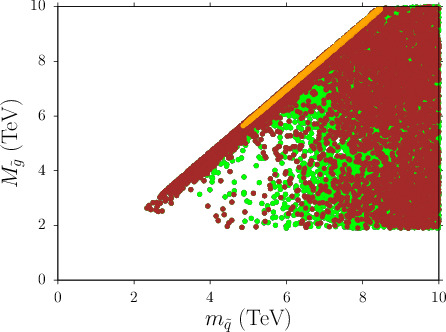}\hfill
        \includegraphics[width=8cm, height=6.5cm]{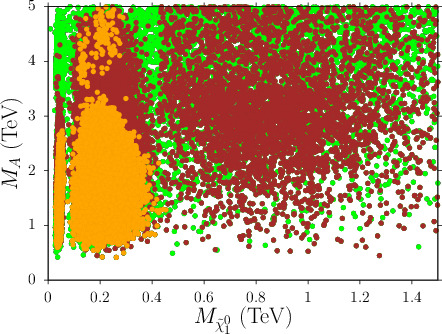}\hfill  
        \includegraphics[width=8cm, height=6.5cm]{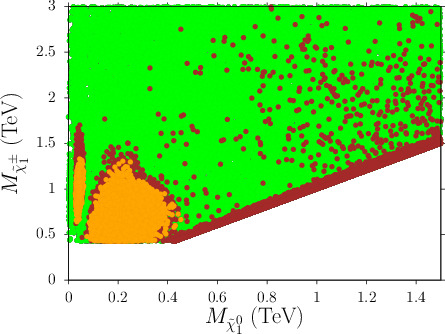}\hfill
         \includegraphics[width=8cm, height=6.5cm]{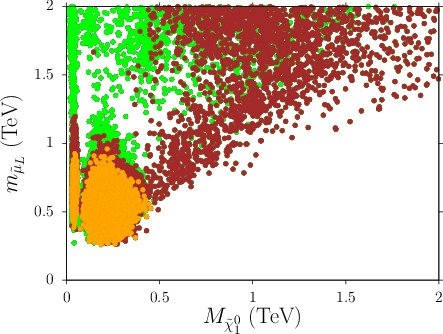}\hfill
         \includegraphics[width=8cm, height=6.5cm]{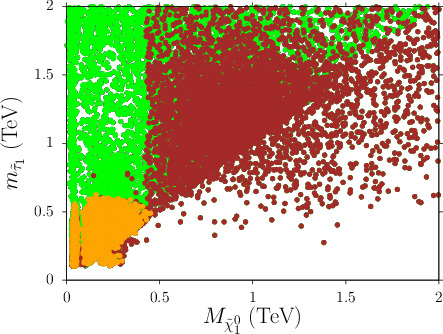}\hfill
         \includegraphics[width=8cm, height=6.5cm]{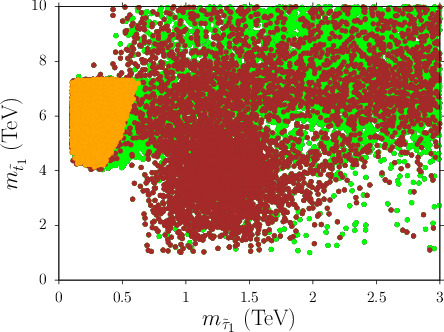}
\caption{Plots in the $m_{\tilde{g}}$ vs. $m_{\tilde{q}}$, $M_A$ vs. $M_{\tilde{\chi}_1^0}$, $M_{\tilde{\chi}_1^\pm}$ vs. $m_{\tilde{\chi}_1^0}$, and $m_{\tilde{t}_1}$ vs. $m_{\tilde{\tau}_1}$ planes. Color coding is the same as Figure \ref{fig:damu-plots}.}
\label{fig:sparticle}
\end{figure}

In Figure \ref{fig:damu-plots2}, our results are displayed in the  $\damu-M_{3}/M_{1}$, $\damu-M_{3}/M_{2}$, $\damu-M_{2}/M_{1}$, $\damu-M_{3}$, $\damu-m_{10}$ and $\damu-m_{\bar{5}}$ planes. We can see from the upper two panels that the gaugino mass ratios are restricted to fairly large values (\textit{orange} points), i.e., $M_{3}/M_{1} \gtrsim 5$ and $M_{3}/M_{2} \gtrsim 2.5$. The ratio $M_{2}/M_{1}$ can be relatively small with a lower bound given by, $M_{2}/M_{1} \gtrsim 1$. From the right central panel, we can see that the gaugino mass parameter $M_3 \gtrsim 3$ TeV if we insist on the resolution of the $\gmu$ anomaly. The large values of the parameter $M_3$ at the GUT scale implies a heavy colored sparticle spectrum at low scale as we can see from the orange points in Figure \ref{fig:sparticle}. From the lower panels of Figure \ref{fig:damu-plots2} we can see that the sfermion masses at $M_{GUT}$ are light and are restricted to fairly narrow intervals, namely,  $250 {\rm \ GeV} \lesssim m_{10} \lesssim 1.25 {\rm \ TeV}$ and $50 {\rm \ GeV} \lesssim m_{\bar{5}} \lesssim 700 {\rm \ GeV}$.

Figure \ref{fig:sparticle} shows our results in the $m_{\tilde{g}}$ vs. $m_{\tilde{q}}$, $M_A$ vs. $M_{\tilde{\chi}_1^0}$, $M_{\tilde{\chi}_1^\pm}$ vs. $m_{\tilde{\chi}_1^0}$,  $m_{\tilde{\mu}_L}$ vs. $M_{\tilde{\chi}_1^0}$, $m_{\tilde{\tau}_1}$ vs. $M_{\tilde{\chi}_1^0}$ and $m_{\tilde{t}_1}$ vs. $m_{\tilde{\tau}_1}$ planes. We can see that the $\gmu$ constraint implies heavy colored sparticle masses bounded in narrow intervals, namely, 
$5 {\rm \ TeV} \lesssim m_{\tilde{q}} \lesssim 8.25 {\rm \ TeV}$,
$5.5 {\rm \ TeV} \lesssim m_{\tilde{g}} \lesssim 10 {\rm \ TeV}$,
$4 {\rm \ TeV} \lesssim m_{\tilde{t}_1} \lesssim 7 {\rm \ TeV}$.
Recent searches for SUSY signatures at the LHC lead to a lower bound of $\sim$ 1.9 TeV on the gluino mass. The High Luminosity LHC (HL-LHC), with an anticipated luminosity of 3000 fb$^{-1}$, will be able to probe gluinos upto 2.3 TeV and stop masses upto 1.2 TeV ~\cite{upgrade,gershtein,atlaswiki}. There is a considerable region of the parameter space of this model that is accessible at the current and future colliders. However, the heavy colored sparticle mass spectrum predicted by the $\gmu$ constraint may be difficult to test at the LHC but might be accessible to future colliders such as the HE-LHC \cite{Baer:2017pba}. We can see from the  $M_A$ vs. $M_{\tilde{\chi}_1^0}$ that the pseudoscalar Higgs boson mass can be as light 500 GeV, which is within the reach of the LHC. In the lower right panel, we can see that the stau can also be light and its mass is bounded in the interval $100 {\rm \ GeV} \lesssim m_{\tilde{\tau}_1} \lesssim 600 {\rm \ GeV}$.

\begin{figure}
\begin{center}
        \includegraphics[width=8cm, height=6.5cm]{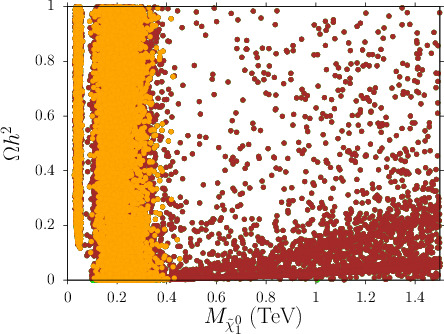}
\end{center}
\caption{Plot in the  $\Omega h^2$ vs. $M_{\tilde{\chi}_1^0}$  plane. Color coding is the same as in Figure \ref{fig:damu-plots}.}
\label{fig:rd}
\end{figure}

In Figure \ref{fig:rd}, we present the  $\Omega h^2$ vs. $M_{\tilde{\chi}_1^0}$ plane. We can see that the parameter space consistent with the $\gmu$ constraint also leads to small relic density for dark matter. We can see from Figure \ref{fig:sparticle} that there are several coannihilation channels such as the chargino-neutralino, smuon-neutralino and stau-neutralino coannihilation channels that come into play in order to yield the desired relic abundance.

\begin{figure}
        \includegraphics[width=8cm, height=6.5cm]{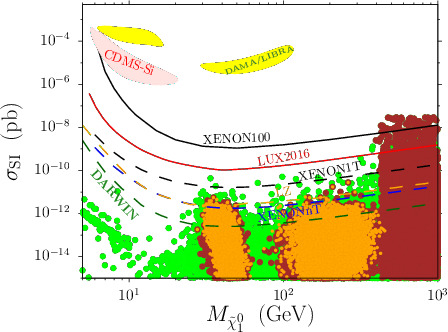}\hfill
        \includegraphics[width=8cm, height=6.5cm]{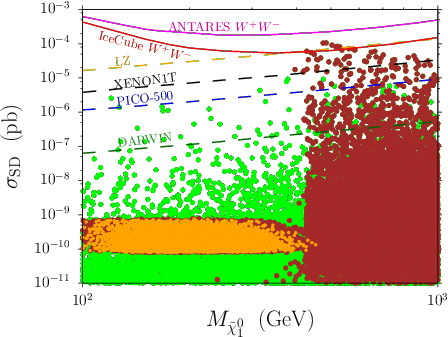} \\ 
\caption{Plots in the  $\sigma_{SI}$ vs. $M_{\tilde{\chi}_1^0}$ and $\sigma_{SD}$ vs. $M_{\tilde{\chi}_1^0}$ planes. Color coding is the same as in Figure \ref{fig:damu-plots}.}
\label{fig:sisd}
\end{figure}

In Figure \ref{fig:sisd}, we analyze the prospects of direct detection of neutralino dark matter in the $\sigma_{SI}$ vs. $M_{\tilde{\chi}_1^0}$ plane. The upper left corner of the plot shows the two anomalous signals DAMA/LIBRA \cite{Savage:2008er} and CDMS-Si \cite{Agnese:2013rvf}. In addition, we display the XENON100 \cite{Aprile:2016swn} and the LUX2016 bound~\cite{Akerib:2016vxi} (solid lines). The future projected reaches of XENON1T~\cite{Aprile:2015uzo}, LZ (with 1 keV cutoff)~\cite{Akerib:2015cja}, XENONnT~\cite{Aprile:2015uzo} and DARWIN ~\cite{Aalbers:2016jon} are shown as dashed lines. We can observe that the parameter space of this model can be probed by direct detection experiments as well. Some of the parameter space is already excluded by the Xenon and LUX experiment (solid lines) whereas a significant region is accessible to the XENON1T, LZ, XENONnT and DARWIN experiments. In particular, the parameter space consistent with the $\gmu$ constraint correspond to low cross sections and will be accessible to the LZ, XENONnT and DARWIN experiments. There is also a notable region of the parameter space with considerably low cross sections which is not accessible to any of the projected sensitivities.

The spin dependent neutralino cross section is displayed in the right panel of Figure \ref{fig:sisd} in the $\sigma_{SD}$ vs. $M_{\tilde{\chi}_1^0}$ plane. The recent limits from Antares ~\cite{Adrian-Martinez:2016gti} and IceCube~\cite{Aartsen:2016exj} are shown as solid lines.  The dashed lines show the projected reach of the LZ~\cite{Akerib:2015cja}, XENON1T~\cite{Aalbers:2016jon}, Pico-500~\cite{ckrauss} and the
 DARWIN~\cite{Aalbers:2016jon} experiments. We can see that the parameter space of this model is  accessible to these future experiments. However the parameter space corresponding to the $\gmu$ constraint (\textit{orange} points) have very low cross sections and is well beyond the search limit of all of these experiments.

\begin{figure}
\begin{center}
        \includegraphics[width=10cm, height=8cm]{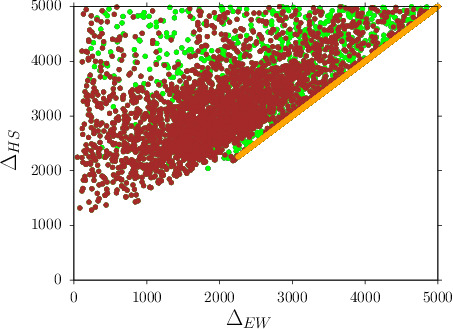}
\end{center}
\caption{Plots in the  $\Delta_{EW}$ vs. $\Delta_{HS}$ planes. Color coding is the same as in Figure \ref{fig:damu-plots}.}
\label{fig:finetuning}
\end{figure}

The version of  Isajet \cite{ISAJET} we employ calculates the  fine-tuning parameters related to the little hierarchy problem at  the Electro Weak ($\Delta_{EW}$)
and GUT scale ($\Delta_{HS}$). In the context of this problem, Figure \ref{fig:finetuning} presents the prediction of this model in the $\Delta_{EW}$ vs. $\Delta_{HS}$ plane. The reader is referred to \cite{Baer:2012mv} for the definition of these parameters. Lower values of these parameters imply that the model is more ``natural'' or less fine tuned. 
 We can see that this model allows for $\Delta_{EW}$ and $\Delta_{HS}$  $\gtrsim 1000$ (fine tuning $< 0.1\%$), which is not very suitable for the resolution of the little hierarchy problem. The $\gmu$ constraint (\textit{orange} points) does not favor the resolution of the little hierarchy problem since it implies $\Delta_{EW} \simeq \Delta_{HS} \gtrsim 2000$  (fine tuning $< 0.05\%$). 

Lastly, we present three benchmark points from our analysis in Table \ref{tab1}. The three points satisfy the various constraints described in Section \ref{sec:parameter}. In addition, the $\gmu$ constraint is also satisfied and the relic density is consistent with WMAP. The neutralino is essentially a bino for the three points. The sleptons are fairly light and the points exhibit stau and smuon coannihilation. The colored sparticle spectrum is quite heavy for the three points ($>$ 6 TeV) and are beyond the reach of LHC. As noted in Figure \ref{fig:finetuning} the parameter space with good $\gmu$ is fine-tuned and we can see that the fine-tuning parameters in the table are $\sim$ 5000 ($0.02 \%$ fine-tuning).

\section{Conclusion}\label{sec:conclude}

We analyzed the supersymmetric $SU(5)$ model with non-universal gaugino masses such that the gaugino mass parameters are independent at the GUT scale. We showed that there is a considerable region of the parameter space of this model that satisfies the sparticle mass constraints, $B$-physics constraints and yields the desired relic abundance. We also showed that the observed deviation in the muon anomalous magnetic moment $\gmu$ can be explained in this model. The sparticle spectroscopy of the model was also presented and it was shown that the colored sparticle spectrum consistent with the $\gmu$ constraint is quite heavy and hardly within the reach of the LHC whereas the sleptons are fairly light. In addition, we analyzed the prospects of direct detection of dark matter in this model and found that the parameter space corresponding to the $\gmu$ constraint predicts low cross sections and is within the projected sensitivites of some experiments. 

\section{Acknowledgments}

The author would like to thank Fariha Nasir for useful discussions.

\newpage

\begin{table}[h!]\vspace{1.5cm}
\centering
\begin{tabular}{|p{3cm}|p{3cm}p{3cm}p{3cm}|}
\hline
\hline
                 	&	 Point 1 	&	 Point 2 	&	 Point 3 	\\
\hline

$m_{10} $         	&$	   315.3	$&$	  460.3	$&$	   537.1	$\\
$m_{\bar{5}} $         	&$	   289	$&$	   413.3	$&$	   312.1	$\\
$M_1$         	&$	397.3	$&$	573.6	$&$	686.5	$\\
$M_2$         	&$	1304.8	$&$	784.8	$&$	680	$\\
$M_3$         	&$	-4448	$&$	-4503	$&$	-4748.7	$\\
$A_0$         	&$	666.4	$&$	996	$&$	878.6	$\\
$\tan\beta$      	&$	36.9	$&$	38.7	$&$	40.6	$\\
\hline		  		  		  	
$\mu$            	&$	901	$&$	715	$&$	678	$\\

\hline		  		  		  	
$m_h$            	&$	124.2	$&$	124.3	$&$	124.5	$\\
$m_H$            	&$	2711	$&$	2417	$&$	2050	$\\
$m_A$            	&$	2693	$&$	2401	$&$	2036	$\\
$m_{H^{\pm}}$    	&$	2712	$&$	2419	$&$	2052	$\\
		  		  		  	
\hline		  		  		  	
$m_{\tilde{\chi}^0_{1,2}}$	&$	  \textbf{209},  1210	$&$	  \textbf{288},   767	$&$	  \textbf{341},   681	$\\

$m_{\tilde{\chi}^0_{3,4}}$	&$	 4459,  4459	$&$	 4605,  4606	$&$	 4847,  4847	$\\

$m_{\tilde{\chi}^{\pm}_{1,2}}$	&$	 1215,  4417	$&$	  769,  4563	$&$	  683,  4802	$\\

$m_{\tilde{g}}$  	&$	8835	$&$	8964	$&$	9427	$\\
		  		  		  	
\hline $m_{ \tilde{u}_{L,R}}$	&$	 7553,  7528	$&$	 7643,  7653	$&$	 8031,  8049	$\\
                 		  		  		  	
$m_{\tilde{t}_{L,R}}$	&$	 6515,  6793	$&$	 6608,  6814	$&$	 6962,  7124	$\\
                 		  		  		  	
\hline $m_{ \tilde{d}_{L,R}}$	&$	 7553,  7532	$&$	 7643,  7655	$&$	 8032,  8040	$\\
                 		  		  		  	
$m_{\tilde{b}_{R}}$	&$	 6737,  6879	$&$	 6751,  6898	$&$	 7040,  7174	$\\
                 		  		  		  	
\hline		  		  		  	
$m_{ \tilde{\mu}_{L,R}}$	&$	  863,   \textbf{276}	$&$	  609,   \textbf{472}	$&$	  \textbf{469},   564	$\\

$m_{\tilde{\tau}_{L,R}}$	&$	  \textbf{243},   942	$&$	  \textbf{328},   845	$&$	  \textbf{382},   913	$\\
                		  		  		  	
\hline		  		  		  	
$\Delta(g-2)_{\mu}$  	&$	  \mathbf{20.6\times 10^{-10}}	$&$	 \mathbf{ 23.5\times 10^{-10}}	$&$	  \mathbf{29.7\times 10^{-10}}	$\\

$\sigma_{SI}({\rm pb})$	&$	  8.71\times 10^{-15}	$&$	  8.63\times 10^{-15}	$&$	  1.15\times 10^{-14}	$\\

$\sigma_{SD}({\rm pb})$	&$	  1.5\times 10^{-10}	$&$	  1.29\times 10^{-10}	$&$	  1.04\times 10^{-10}	$\\

$\Omega_{CDM}h^{2}$	&$	0.12	$&$	0.12	$&$	0.1	$\\

$\Delta_{EW}$	&$	4945	$&$	5278	$&$	5852	$\\

$\Delta_{HS}$	&$	4945	$&$	5278	$&$	5852	$\\

\hline

\end{tabular}
\caption{Masses in the table are in units of GeV. All the points satisfy the $B$-physics and sparticle mass constraints presented in section \ref{sec:parameter}. In addition the points satisfy the $\gmu$ constraint. The smuons and staus are fairly light for the three points and these points exhibit smuon and stau coannihilation.  }
\label{tab1}
\end{table}


\newpage


\newpage


\begin{thebibliography}{99}

\bibitem{Jungman:1995df}
  See, for instance, G.~Jungman, M.~Kamionkowski and K.~Griest,
  Phys.\ Rept.\  {\bf 267}, 195 (1996).


%
\bibitem{lhc-squark} ATLAS collaboration, ATLAS-CONF-2017-022; CMS
  Collaboration, CMS-SUS-16-036.
%
\bibitem{lhc-stop} ATLAS Collboration, ATLAS-CONF-2017-020; CMS
  Collboration, CMS-SUS-16-051 and CMS-SUS-16-049.

\bibitem{lhc-chargino} ATLAS Collboration, ATLAS-CONF-2017-017.
%
\bibitem{upgrade} See, {\it e.g.} ATLAS Phys. PUB 2013-011; CMS
  Note-13-002.
%
\bibitem{gershtein}
Y.~Gershtein {\it et al.},
arXiv:1311.0299 [hep-ex].

%
\bibitem{atlaswiki}
\verb^https://twiki/cern.ch/bin/view/AtlasPublic/UpgradePhysicsStudies^

\bibitem{Baer:2017pba}
  See, for instance,
  A.~Aboubrahim and P.~Nath,
  Phys.\ Rev.\ D {\bf 96}, 075015 (2017);
  H.~Baer, V.~Barger, J.~S.~Gainer, H.~Serce and X.~Tata,
  arXiv:1708.09054 [hep-ph];
  K.~Kowalska, L.~Roszkowski, E.~M.~Sessolo and A.~J.~Williams,
  JHEP {\bf 1506}, 020 (2015);
  C.~Han, K.~i.~Hikasa, L.~Wu, J.~M.~Yang and Y.~Zhang,
  Phys.\ Lett.\ B {\bf 769}, 470 (2017);
  W.~Ahmed, X.~J.~Bi, T.~Li, J.~S.~Niu, S.~Raza, Q.~F.~Xiang and P.~F.~Yin,
  arXiv:1709.06371 [hep-ph];
  J.~Kawamura and Y.~Omura,
  JHEP {\bf 1708}, 072 (2017)
  doi:10.1007/JHEP08(2017)072
  [arXiv:1703.10379 [hep-ph]];
  J.~Kawamura and Y.~Omura,
  Phys.\ Rev.\ D {\bf 93}, no. 5, 055019 (2016)
  doi:10.1103/PhysRevD.93.055019
  [arXiv:1601.03484 [hep-ph]];
  H.~Abe, J.~Kawamura and Y.~Omura,
  JHEP {\bf 1508}, 089 (2015)
  doi:10.1007/JHEP08(2015)089
  [arXiv:1505.03729 [hep-ph]].

\bibitem{Hagiwara:2011af}
  M.~Davier, A.~Hoecker, B.~Malaescu and Z.~Zhang,
  Eur.\ Phys.\ J.\ C {\bf 71}, 1515 (2011)
  [Erratum-ibid.\ C {\bf 72}, 1874 (2012)];
  K.~Hagiwara, R.~Liao, A.~D.~Martin, D.~Nomura and T.~Teubner,
  J.\ Phys.\ G {\bf 38}, 085003 (2011).


\bibitem{Moroi:1995yh}
  T.~Moroi,
  Phys.\ Rev.\ D {\bf 53}, 6565 (1996)
  [Erratum-ibid.\ D {\bf 56}, 4424 (1997)];
  
  
\bibitem{Martin:2001st}
  S.~P.~Martin and J.~D.~Wells,
  Phys.\ Rev.\ D {\bf 64}, 035003 (2001).
  G.~F.~Giudice, P.~Paradisi and A.~Strumia,
  JHEP {\bf 1210}, 186 (2012)
 
\bibitem{Profumo:2003ema}
  S.~Profumo,
  Phys.\ Rev.\  D {\bf 68}, 015006 (2003);
  B.~Ananthanarayan and P.~N.~Pandita,
  Int.\ J.\ Mod.\ Phys.\  A {\bf 22}, 3229 (2007).

\bibitem{Gogoladze:2008dk}
  I.~Gogoladze, R.~Khalid, N.~Okada and Q.~Shafi,
  Phys.\ Rev.\ D {\bf 79}, 095022 (2009);
  H.~Baer, I.~Gogoladze, A.~Mustafayev, S.~Raza and Q.~Shafi,
  JHEP {\bf 1203}, 047 (2012).


\bibitem{Kawamura:2017amp} 
  J.~Kawamura and Y.~Omura,
  JHEP {\bf 1708}, 072 (2017)
  doi:10.1007/JHEP08(2017)072
  [arXiv:1703.10379 [hep-ph]];
  H.~Abe, J.~Kawamura and Y.~Omura,
  JHEP {\bf 1508}, 089 (2015)
  doi:10.1007/JHEP08(2015)089
  [arXiv:1505.03729 [hep-ph]];
  J.~Kawamura and Y.~Omura,
  Phys.\ Rev.\ D {\bf 93}, no. 5, 055019 (2016)
  doi:10.1103/PhysRevD.93.055019
  [arXiv:1601.03484 [hep-ph]];
  S.~P.~Martin,
  Phys.\ Rev.\ D {\bf 89}, no. 3, 035011 (2014)
  doi:10.1103/PhysRevD.89.035011
  [arXiv:1312.0582 [hep-ph]];
  M.~Badziak, M.~Olechowski and S.~Pokorski,
  JHEP {\bf 1310}, 088 (2013)
  doi:10.1007/JHEP10(2013)088
  [arXiv:1307.7999 [hep-ph]];
  S.~Caron, J.~Laamanen, I.~Niessen and A.~Strubig,
  JHEP {\bf 1206}, 008 (2012)
  doi:10.1007/JHEP06(2012)008
  [arXiv:1202.5288 [hep-ph]].


\bibitem{Akula:2013ioa}
  M.~Endo, K.~Hamaguchi, S.~Iwamoto and T.~Yoshinaga,
  JHEP {\bf 1401}, 123 (2014);
  S.~Mohanty, S.~Rao and D.~P.~Roy,
  JHEP {\bf 1309}, 027 (2013);
  S.~Akula and P.~Nath,
  Phys.\ Rev.\ D {\bf 87}, 115022 (2013);
  M.~Endo, K.~Hamaguchi, T.~Kitahara and T.~Yoshinaga,
  JHEP {\bf 1311}, 013 (2013);
  J.~Chakrabortty, S.~Mohanty and S.~Rao,
  arXiv:1310.3620 [hep-ph].
  
  
\bibitem{Gogoladze:2014cha} 
  I.~Gogoladze, F.~Nasir, Q.~Shafi and C.~S.~Un,
  Phys.\ Rev.\ D {\bf 90}, no. 3, 035008 (2014)
  doi:10.1103/PhysRevD.90.035008
  [arXiv:1403.2337 [hep-ph]].


\bibitem{Gogoladze:2016jvm} 
  I.~Gogoladze and C.~S.~Un,
  Phys.\ Rev.\ D {\bf 95}, no. 3, 035028 (2017)
  doi:10.1103/PhysRevD.95.035028
  [arXiv:1612.02376 [hep-ph]];
  I.~Gogoladze, Q.~Shafi and C.~S.~Ün,
  Phys.\ Rev.\ D {\bf 92}, no. 11, 115014 (2015)
  doi:10.1103/PhysRevD.92.115014
  [arXiv:1509.07906 [hep-ph]];
  J.~Chakrabortty, A.~Choudhury and S.~Mondal,
  JHEP {\bf 1507}, 038 (2015)
  doi:10.1007/JHEP07(2015)038
  [arXiv:1503.08703 [hep-ph]].
  
 
\bibitem{Ajaib:2014ana}
  M.~A.~Ajaib, I.~Gogoladze, Q.~Shafi and C.~S.~Un,
  arXiv:1402.4918 [hep-ph].
  
\bibitem{Martin:2009ad}
 See, for instance,  S.~P.~Martin,
  Phys.\ Rev.\  {\bf D79}, 095019 (2009);
    U.~Chattopadhyay, D.~Das and D.~P.~Roy,
  Phys.\ Rev.\  D {\bf 79}, 095013 (2009);
   B.~Ananthanarayan, P.~N.~Pandita,
  Int.\ J.\ Mod.\ Phys.\  {\bf A22}, 3229-3259 (2007);
  S.~Bhattacharya, A.~Datta and B.~Mukhopadhyaya,
  JHEP {\bf 0710}, 080 (2007);
  A.~Corsetti and P.~Nath,
  Phys.\ Rev.\  D {\bf 64}, 125010 (2001)
   and references therein.
 
\bibitem{Martin:2013aha}
  S.~P.~Martin,
  arXiv:1312.0582 [hep-ph].

\bibitem{Anandakrishnan:2013cwa}
  A.~Anandakrishnan and S.~Raby,
  Phys.\ Rev.\ Lett.\  {\bf 111}, 211801 (2013).
  






  
\bibitem{ISAJET}
  F.~E.~Paige, S.~D.~Protopopescu, H.~Baer and X.~Tata,
  hep-ph/0312045.

\bibitem{Belanger:2008sj}
  G.~Belanger, F.~Boudjema, A.~Pukhov and A.~Semenov,
  Comput.\ Phys.\ Commun.\  {\bf 180}, 747 (2009).


\bibitem{Leva}
J.L. Leva,
 Math. Softw. 18 (1992) 449;
J.L. Leva,
Math. Softw. 18 (1992) 454.




\bibitem{Ajaib:2015ika} 
  M.~Adeel Ajaib, I.~Gogoladze and Q.~Shafi,
  Phys.\ Rev.\ D {\bf 91}, no. 9, 095005 (2015)
  doi:10.1103/PhysRevD.91.095005
  [arXiv:1501.04125 [hep-ph]].
 





\bibitem{Pokoroski}
  M.~Badziak, Z.~Lalak, M.~Lewicki, M.~Olechowski and S.~Pokorski,
  JHEP {\bf 1503}, 003 (2015)
  [arXiv:1411.1450 [hep-ph]].


  

  
\bibitem{Savage:2008er} 
  C.~Savage, G.~Gelmini, P.~Gondolo and K.~Freese,
  JCAP {\bf 0904}, 010 (2009)
  doi:10.1088/1475-7516/2009/04/010
  [arXiv:0808.3607 [astro-ph]].
  

  
\bibitem{Agnese:2013rvf} 
  R.~Agnese {\it et al.} [CDMS Collaboration],
  Phys.\ Rev.\ Lett.\  {\bf 111}, no. 25, 251301 (2013)
  doi:10.1103/PhysRevLett.111.251301
  [arXiv:1304.4279 [hep-ex]].
  
%
\bibitem{Aprile:2016swn}
  E.~Aprile {\it et al.} [XENON100 Collaboration],
  arXiv:1609.06154 [astro-ph.CO].
  
\bibitem{Akerib:2016vxi}
  D.~S.~Akerib {\it et al.},
  arXiv:1608.07648 [astro-ph.CO].
  
  
%
\bibitem{Aprile:2015uzo}
  E.~Aprile {\it et al.} [XENON Collaboration],
  JCAP {\bf 1604} (2016) no.04,  027
  doi:10.1088/1475-7516/2016/04/027
  [arXiv:1512.07501 [physics.ins-det]].
  
%
\bibitem{Akerib:2015cja}
  D.~S.~Akerib {\it et al.} [LZ Collaboration],
  arXiv:1509.02910 [physics.ins-det].






%
\bibitem{Aalbers:2016jon}
  J.~Aalbers {\it et al.} [DARWIN Collaboration],
  arXiv:1606.07001 [astro-ph.IM].
%



%
\bibitem{Adrian-Martinez:2016gti}
  S.~Adrian-Martinez {\it et al.} [ANTARES Collaboration],
  Phys.\ Lett.\ B {\bf 759} (2016) 69
  doi:10.1016/j.physletb.2016.05.019
  [arXiv:1603.02228 [astro-ph.HE]].
%

%
\bibitem{Aartsen:2016exj}
  M.~G.~Aartsen {\it et al.} [IceCube Collaboration],
  JCAP {\bf 1604} (2016) no.04,  022
  doi:10.1088/1475-7516/2016/04/022
  [arXiv:1601.00653 [hep-ph]].
%
\bibitem{ckrauss} Talk by C. Krauss for the Pico collaboration, ICHEP 2016 meeting, Chicago, IL, August 2016.


\bibitem{Baer:2012mv}
  H.~Baer, V.~Barger, P.~Huang, D.~Mickelson, A.~Mustafayev and X.~Tata,
  arXiv:1210.3019 [hep-ph].



\bibitem{Gogoladze:2012yf}
  I.~Gogoladze, F.~Nasir and Q.~Shafi,
  arXiv:1212.2593 [hep-ph].


 

\end{thebibliography}
\end{document}